\input{aipcheck}

\documentclass[
    ,final            
  ]
  {aipproc}

\layoutstyle{8x11double}

\begin{document}

\title{Ageing of a granular pile induced by thermal cycling}


\author{Thibaut Divoux}{}

\author{Ion Vassilief}{}

\author{Herv\'e Gayvallet}{}

\author{Jean-Christophe G\'eminard}{
  address={Universit\'e de Lyon, Laboratoire de Physique, Ecole Normale Sup\'erieure de Lyon, CNRS, 46 All\'ee d'Italie, 69364 Lyon cedex 07, France}
}

\begin{abstract}
Here we show that variations of temperature, even of a few degrees in amplitude, induce the ageing of a granular pile.
In particular, we report measurements of physical properties of a granular heap submitted to thermal cycles. Namely, we focus on the evolution of the thermal linear-expansion coefficient and of the thermal conductivity of the pile with the number of cycles.
The present contribution nicely supplements a recent article we published elsewhere [Phys. Rev. Lett. \textbf{101}, 148303 (2008)] and introduces a different and promising method to impose temperature cycles to a granular pile.
\end{abstract}

\maketitle


\section{Introduction}
A static granular pile is essentially a {\it fragile} construction \cite{Cates,Roux}; Minute perturbations applied at the microscopic scale (the surface roughness of a grain) can lead to macroscopic reorganisations of the pile \cite{Claudin}. A remarkable signature of this fragility is the sensitivity of a granular pile to temperature variations. First presented as a hindrance to assess the sound propagation in sand \cite{Liu1992} or to perform reproducible measurements of the stress distribution at the base of a granular pile \cite{Clement}, the sensitivity of granular matter to temperature fluctuations has been since used to pack grains \cite{Geminard03,Chen,Slotterback}. Indeed, a granular heap submitted to temperature cycles, even of a few degrees in amplitude \cite{Bonamy,Geminard03,Djaoui,Divoux08}, experiences successive large-scale "static avalanches" \cite{Claudin}, which induces the slow compaction of this fragile construction. Such reorganisations are possible because of the stress anisotropy inside the pile \cite{Travers} and of the surface properties of the grains \cite{Bonamy, Divoux08}.

In a recent letter \cite{Divoux08}, we reported a time-resolved study of the dynamics associated with the compaction of a granular column submitted to thermal cycles.
Here, we demonstrate that temperature variations, even of a few degrees in amplitude, also induce ageing in a granular pile exactly as moisture \cite{Bocquet}, constant applied-stress \cite{Losert} and chemical reactions between grains \cite{Gayvallet} do.  Here, we dwell on two physical properties of the granular pile, namely $\kappa_g$, its thermal linear-expansion coefficient, and $\lambda_g$, its thermal conductivity which both evolve in time because of temperature varations. Both physical properties remain poorly studied in the case of a 3D pile free to dilate and, thus, to reorganise under temperature variations. Indeed, previously, only $\lambda_g$ has been assessed numerically and experimentaly in the case of a bidimensional and compressed static-bed (see \cite{Vargas01} and references therein). We thus address here the issue of the dependence of $\kappa_b$ and $\lambda_b$ with the numbers of imposed temperature-cycles.

\begin{figure}[!h]
\includegraphics[width=0.9\columnwidth]{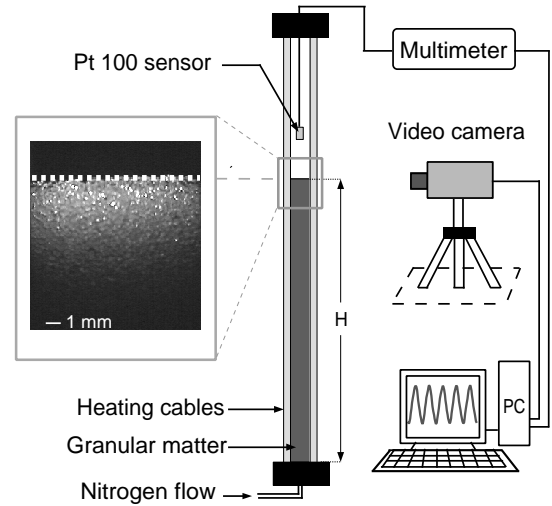}
\caption{\small{{\bf Sketch of the first experimental setup.} Inset: picture of the upper part
of the column. The granular level is indicated by the white dotted-line.}
}
\label{fig.1}
\end{figure}

\section{Thermal dilation}

In this section, we report, in a first experimental configuration, the evolution of the thermal linear-expansion coefficient of a granular column submitted to thermal cycles. The experiment consists in imposing temperature cycles to a granular column and in measuring the resulting variations of the column height.
 
\subsection{Experimental setup}

The experimental set-up (Fig.~\ref{fig.1}) consists in a vertical glass tube (height 1.7~m, inner diameter 13~mm) firmly fastened to a wall into the basement of the physics department, in order to flee mechanical vibrations. The sample consists of a column (height $H$) of spherical glass beads (diameter $d$) poured into the tube. An additional gas-input, at the bottom of the column, makes possible to loosen the pile thanks to an upward flow of dry nitrogen.
The temperature cycles are imposed by means of a heating cable (Prolabo, 40~W/m) directly taped on
the outer surface of the tube wall. The resulting temperature is measured by means
of a sensor (Pt100, located close to the free surface of the granular material)
and a multimeter (Keithley, 196). 
The free surface of the material, which is illuminated by a red LED (Kingbright, L-793SRC-E,
located inside the tube, above the granular material) is imaged from the side with a video camera (Panasonic, WV-BP500) connected to a frame grabber board (Data Translation, DT2255).
A macro, running under a data-processing software (WaveMetrics, IGOR Pro 4.0),
drives the heating power, records the resulting variations of the temperature
and measures accurately the height $H$ from the images:
A subpixel resolution (namely, less than a tenth of a pixel which typically stands for 5 $\mu $m)
is achieved by considering the average position of the free surface, assumed to correspond to the
inflection point in the vertical intensity-profile averaged over the whole diameter of the tube.
Measurements are performed $20$ times per temperature cycle.

Due to long experimental times, we limit our report to a given diameter
$d = (510 \pm 90)~\mu$m of the grains (Matrasur Corp.) and to a given period $2\pi/\omega = 600$~s of the cycles. The cycling period, 10 minutes, is arbitrarily chosen to be small enough to avoid excessively-long experimental times but large enough to insure that the associated thermal penetration-length $l_p\equiv \sqrt{ 2\lambda/(\, C \omega)} \simeq 6$~mm is about the tube radius
($\lambda \simeq$~0.2~W m$^{-1}$ K$^{-1}$ and $C \simeq 10^6$~J m$^{-3}$ K$^{-1}$ respectively
denote the thermal conductivity and heat capacity of a typical glass-grains pile \cite{Geminard01}.)
It is here crucial to note that the column is heated homogeneously along its whole length but
that the temperature is likely to vary in the radial direction.

Prior to each experiment, the granular column is prepared in a low-density state thanks to the dry-nitrogen upward flow. The top of the column is then higher than the field imaged by the camera (typically 1~cm above) and we set the amplitude of the cycles, $\Delta T$, to the largest accessible value, $\Delta T=27.1^{\circ}$ C.
The preparation of the sample ends when the top of the column enters the observation field.
At this point, the granular column is "quenched": The amplitude of the cycles is set to the
chosen value $\Delta T$ lying between $10.8$ and $27.1 ^{\circ}$C, which defines the origin of time $t= 0$. The granular column is subsequently submitted to, at least, $1000$ cycles (7 days).

\begin{figure}[!t]
\includegraphics[width=0.9\columnwidth]{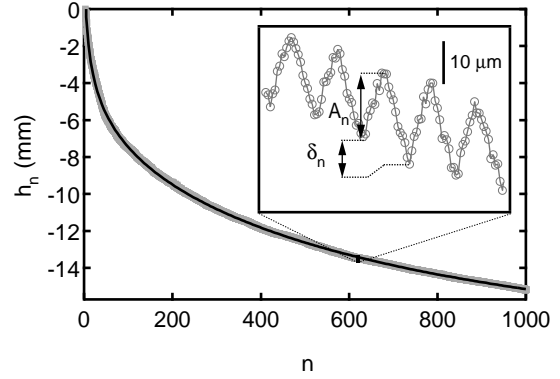}
\caption{\small{\bf Height variation $h_n$ vs. number of cycles $n$.}
One observes first an exponential behavior at short time followed by a subsequent
logarithmic creep at long time [The black curve corresponds to the test function $h_n^t\equiv\,h_0+h_e\,\exp{(-n/n_c)} + h_l\,\ln(n)$.]
Inset - Oscillations of the column height associated with the temperature cycles :
$A_n$ and $\delta _n$ are respectively defined to be the amplitude of the increase and the drift of $h_n$
at the cycle $n$
($H = 140$~cm, $2 \pi/\omega = 600$~s and $\Delta T=10.8^{\circ}$C.)
}
\label{fig.2}
\end{figure}

\subsection{Results}

Under the action of the temperature cycles the column height decreases:
we report the variation $h_n\equiv~H(2\pi~n/\omega)-H(0)$,
where $n$ denotes the time in units of the cycle period or, equivalently when integer, the number
of imposed cycles (Fig.~\ref{fig.2}).
We observe that the thermal-induced compaction is a very slow phenomenon:
after 7 days ($1000$ cycles), the decrease of the height is of about 1.5~cm
(about 1\% of the height $H$), which indicates that the system remains very far
from the maximum compaction (roughly a decrease of about 10\% of the column height, i.e. $h_n \sim 10$~cm) within the experimental time.
Accordingly, after the experiment, we checked that a single finger tap produces a
collapse of the granular height of one centimeter at a rough estimate.
We also checked during 3 days that the height $H$ of the column stays constant when no
temperature cycles are imposed, which proves that ambient mechanical vibrations
and changes in the room temperature have no (or little) effect in our experimental
conditions.

\begin{figure}[!t]
\includegraphics[width=0.9\columnwidth]{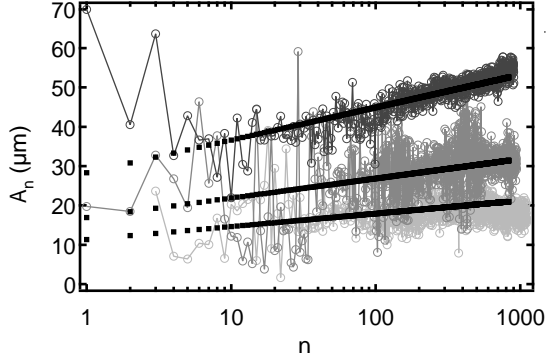}
\caption{\small{\bf Amplitude $A_n$ vs. number of cycles $n$.}
Black lines : $A_n=\Delta T \, [a_0+b_0$ ln$(n)]$
($H = 140$~cm, $2 \pi/\omega = 600$~s and, from bottom to top,
$\Delta T=10.8, 16.2 {\rm~and~} 27.1^{\circ}$C.)}
\label{fig.dilation}
\end{figure}

Interestingly, the measurements are accurate enough to reveal
the oscillations of $H$ associated with the temperature variations (Fig.~\ref{fig.2}, inset).
We observe on the raw data that the amplitude $A_n$, which is proportional to
$\Delta T$, increases logarithmically with $n$ (Fig.~\ref{fig.dilation}).
The oscillations of $H$ are due to the thermal dilation of both the tube
and the granular material. In order to assess the contribution of the granular
material alone, we first determine the amplitude,
$\delta h_t(z)$, of the tube displacement in the laboratory frame as a function
of the height $z$ (origin at the bottom of the column) by marking its outer wall.
The amplitude $\delta h_t(z)$ is found to be linear in $z$ and
the slope provides us with an estimate of the linear thermal-expansion coefficient,
$\kappa = (3.6 \pm 0.4)\,\times 10^{-6}$~K$^{-1}$, of the tube material.
Then, considering the relative variation of the inner volume,
we write the relation  between the amplitude $A_n$ and
the relative variation, $\delta V_g/V_g$, of the volume $V_g$
of the granular material:
$A_n - \delta h_t(0) = H (\delta V_g/V_g - 2 \kappa \Delta T)$.
Experiments performed for different height $H$
demonstrate that $A_n - \delta h_t(0)$ is proportional to $H$,
which shows that $\delta V_g/V_g$ is independent of $H$ and, thus,
that the whole height $H$ of the granular column is
involved in the observed oscillations of the free surface.
For a homogeneous temperature-variation, $\Delta T$,
in the whole cross-section of the tube, we would have $\delta V_g/V_g = 3 \kappa_g \Delta T$.
With this assumption, we obtain a rough estimate of the thermal expansion
coefficient of the granular material, $\kappa_g \simeq [ 3.4 + 0.03 \ln(n)]\times 10^{-6}$~K$^{-1}$,
which is thus found to be very close to $\kappa$ and to increase logarithmically by about 5\%
during the first 1000 cycles (Fig.~\ref{fig.dilation}) as a consequence of
a variation of about 1\% of the density (Fig.~\ref{fig.2}).

\section{Thermal conductivity}

In this section, using a second experimental setup, we estimate the thermal conductivity of a granular pile and perform a time resolved study of its evolution under cycles of temperature. The experiment consists in heating and in measuring the temperature of a granular column along its axis of revolution thanks to a thin heating wire.

\begin{figure}[!t]
\includegraphics[width=0.9\columnwidth]{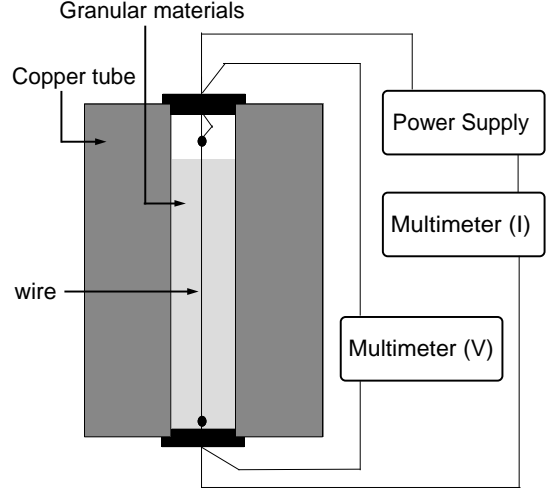}
\caption{\small{\bf Sketch of the second experimental setup.}}
\label{fig.setup2}
\end{figure}

\subsection{Experimental setup}
 
The experimental setup (Fig.~\ref{fig.setup2}) consists in a copper tube (inner diameter 1~cm; outer diameter 5~cm; length 15~cm) partially filled with glass beads (diameter $d\in[250-425]~\mu$m). The tube is maintained at a constant temperature $T_{e}=55^{\circ}$C (numerical PID controller written in C++, precision $0.1^{\circ}$C) whereas the granular column is heated along its axis thanks to a nickel wire (diameter $r_w$~=100~$\mu$m) connected to a power supply (Hewlett Packard, 6633A). We measure simultaneously the imposed current $I$ and the resulting voltage $V$ (Keithley, 196) in order to deduce the heating power $P = U I$ and the wire temperature $T_w$ which we deduce from the resistance $R_w = U/I$. The thermal conductivity $\lambda_g$ of the granular material is then assessed from the temperature difference $T_w-T_e$ which depends linearly on $P/\lambda_g$ \cite{Geminard01}.


\subsection{Results}

In order to measure the thermal conductivity, $P$ is increased by steps (typically 10) up to $T_w-T_e \simeq 40^{\circ}$C and then decreased by steps
down to 0. For each value of $P$, we wait until the steady state is reached and measure the associated voltage $U$ and current $I$.
We obtain $\lambda_g = (0.16 \pm 0.02)$ W/m/K. This value is compatible with the value expected for a pile of glass beads ($\lambda_{glass}=1.4$ W/m/K) surrounded by air ($\lambda_{air}=0.025$ W/m/K) \cite{Geminard01}.

\begin{figure}[!t]
\includegraphics[width=0.9\columnwidth]{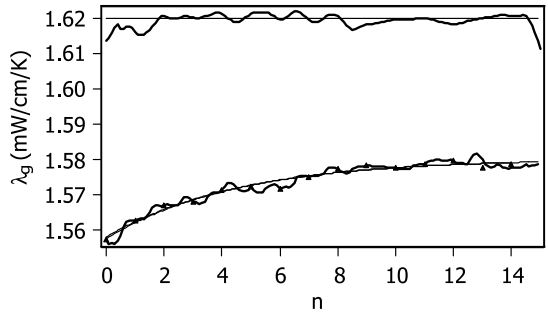}
\caption{\small{\bf Thermal conductivity $\lambda_g$ vs. number of cycles $n$.}
The upper curve corresponds to a pile which has been gently tapped previous to the experiment. The lower curve corresponds to a loose pile. Note first that the conductivity is larger for a larger density. Moreover, the conductivity of the loose material increases significantly when one imposes the thermal cycles ( $d\, \in$ [250-425], $\Delta T=40^{\circ}$C.)}
\label{fig.conductivite}
\end{figure}

We interestingly also observed that, for the same material, $\lambda_g$ slightly depends on the preparation (Fig.~\ref{fig.conductivite})~:
$\lambda_g \simeq 0.162$ W/m/K if the system is tapped previous to the measurement, $\lambda_g \simeq 0.156$ W/m/K if not. It is then particulary interesting to consider the behaviour of the sample when subjected to several temperature cycles.
When the measurements are repeated several times, one observes that the thermal
conductivity of the loose sample significantly increases with the number $n$ of imposed cycles. By contrast, the conductivity of the
tapped sample only slightly fluctuates around a constant value.

The second experimental configuration does not make possible to observe the grains. However, a few experiments reported in details in \cite{Geminard03} proved that the granular column is subjected to compaction even if the container does not dilate (heating cable along the column axis). Our results show that the increase in the thermal conductivity $\lambda_g$ with the number of cycles $n$ originates in the slow compaction of the material.

\section{Conclusion and outlooks}

We reported measurements of the thermal linear-expansion coefficient $\kappa_g$ and of the thermal conductivity $\lambda_g$ of a granular pile.
We have shown that temperature variations, even of a few degrees in amplitude, can lead to the compaction of the pile and, as
consequences, to increases in $\kappa_g$ and $\lambda_g$. Thus, uncontrolled temperature variations can be responsible of part of the
observed ageing of the physical properties of a granular material at rest.

Our results also emphasize how delicate is the thermal cycling method compared to the classic tapping method \cite{Richard}, even applied at low tapping amplitude \cite{Kabla,Umbanhowar}. In this sense, both methods we describe here certainly deserve further study and could be of great interest to probe the jamming transition of a granular assembly \cite{Slotterback} and to unravel the local structure of grain displacement in the vicinity of this transition \cite{Tsamados}.   

The mechanisms at stake in the compaction itself are still not well understood and deserve to be further invertigated. We already mentionned that the dilation of the container is not necessary, which indicates that the difference between the thermal expansion coefficient of the beads and that of the container is probably not the primary cause of the compaction, contrary to what was previously proposed \cite{Chen}. We are currently investigating the compaction of a granular column heated along its axis : A study of the response of the system as a function of the cycling period makes possible to analyze the role played by the penetration length of the temperature field (by the temperature gradient rather than by the temperature variations) in the compaction process.

\end{document}